\documentclass[sigconf]{acmart}

\usepackage{amsfonts}
\usepackage{graphicx}
\usepackage{subcaption}
\usepackage{textcomp}
\usepackage{xcolor}
\usepackage{enumitem}
\usepackage{url}
\usepackage{soul}
\usepackage[]{collab}

\newcommand\alias{\textsc{Cast}\xspace}

\collabAuthor{zb}{teal}{Zhuangbin}

\AtBeginDocument{%
  }

\copyrightyear{2026}
\acmYear{2026}
\setcopyright{cc}
\setcctype{by-nc-nd}
\acmConference[ICSE-SEIP '26]{2026 IEEE/ACM 48th International Conference on Software Engineering}{April 12--18, 2026}{Rio de Janeiro, Brazil}
\acmBooktitle{2026 IEEE/ACM 48th International Conference on Software Engineering (ICSE-SEIP '26), April 12--18, 2026, Rio de Janeiro, Brazil}
\acmPrice{}
\acmDOI{10.1145/3786583.3786867}
\acmISBN{979-8-4007-2426-8/2026/04}

\begin{document}

\title{\alias: Automated Resilience Testing for Production Cloud Service Systems}



\author{Zhuangbin Chen}
\affiliation{%
  \institution{School of Software Engineering\\Sun Yat-sen University}
  \city{Zhuhai}
  \state{Guangdong}
  \country{China}}
\email{chenzhb36@mail.sysu.edu.cn}
\orcid{0000-0001-5158-6716}

\author{Zhiling Deng}
\affiliation{%
  \institution{School of Software Engineering\\Sun Yat-sen University}
  \city{Zhuhai}
  \state{Guangdong}
  \country{China}}
\email{dengzhling3@mail2.sysu.edu.cn}
\orcid{0009-0009-8715-8962}

\author{Kaiming Zhang}
\affiliation{%
  \institution{School of Software Engineering\\Sun Yat-sen University}
  \city{Zhuhai}
  \state{Guangdong}
  \country{China}}
\email{zhangkm7@mail2.sysu.edu.cn}
\orcid{0009-0000-8570-3298}

\author{Yang Liu}
\affiliation{%
  \institution{School of Software Engineering\\Sun Yat-sen University}
  \city{Zhuhai}
  \state{Guangdong}
  \country{China}}
\email{liuy2355@mail2.sysu.edu.cn}
\orcid{0009-0007-0816-0657}


\author{Cheng Cui}
\affiliation{%
  \institution{Huawei Cloud}
  \city{Shenzhen}
  \state{Guangdong}
  \country{China}}
\email{cuicheng2@huawei.com}
\orcid{0009-0006-2558-9335}

\author{Jinfeng Zhong}
\affiliation{%
  \institution{Huawei Cloud}
  \city{Shenzhen}
  \state{Guangdong}
  \country{China}}
\email{zhongjinfeng@huawei.com}
\orcid{0009-0004-6134-7080}

\author{Zibin Zheng}
\authornote{Zibin Zheng is the corresponding author.}
\affiliation{%
  \institution{School of Software Engineering\\Sun Yat-sen University}
  \city{Zhuhai}
  \state{Guangdong}
  \country{China}}
\email{zhzibin@mail.sysu.edu.cn}
\orcid{0000-0001-7872-7718}

\renewcommand{\shortauthors}{Chen et al.}

\begin{abstract}
The distributed nature of microservice architecture introduces significant resilience challenges.
Traditional testing methods, limited by extensive manual effort and oversimplified test environments, fail to capture production system complexity.
To address these limitations, we present \alias, an automated, end-to-end framework for microservice resilience testing in production.
It achieves high test fidelity by replaying production traffic against a comprehensive library of application-level faults to exercise internal error-handling logic.
To manage the combinatorial test space, \alias employs a complexity-driven strategy to systematically prune redundant tests and prioritize high-value tests targeting the most critical service execution paths.
\alias automates the testing lifecycle through a three-phase pipeline (i.e., startup, fault injection, and recovery) and uses a multi-faceted oracle to automatically verify system resilience against nuanced criteria.
Deployed in Huawei Cloud for over eight months, \alias has been adopted by many service teams to proactively address resilience vulnerabilities.
Our analysis on four large-scale applications with millions of traces reveals 137 potential vulnerabilities, with 89 confirmed by developers.
To further quantify its performance, \alias is evaluated on a benchmark set of 48 reproduced bugs, achieving a high coverage of 90\%.
The results show that \alias is a practical and effective solution for systematically improving the reliability of industrial microservice systems.

\end{abstract}




\maketitle
\section{Introduction}
\label{sec: intro}


The widespread adoption of microservice architecture has fundamentally transformed how modern cloud applications are designed and deployed. This architectural paradigm offers significant advantages in terms of agility, scalability, and maintainability by decomposing monolithic applications into loosely coupled, independently deployable services. However, the highly distributed nature of microservice systems introduces substantial challenges regarding \textit{system resilience}, particularly in the face of complex and often non-deterministic failures in cloud environments~\cite{DBLP:conf/sigsoft/ChenKLZZXZYSXDG20,DBLP:journals/csur/WelshB20}.

\textit{Resilience testing} plays a critical role in ensuring the reliability of microservice applications~\cite{DBLP:conf/nsdi/GunawiDJAHAASB10,DBLP:journals/usenix-login/YuanLZRZZJS15,DBLP:conf/sosp/LuL00TYY19}.
Unlike traditional testing approaches that focus on functional correctness~\cite{DBLP:conf/icse/AtlidakisGP19,DBLP:journals/tosem/Arcuri19,DBLP:journals/jss/HuiWLYSZCL25}, resilience testing evaluates a software system's ability to tolerate, adapt to, and recover gracefully from failures once the underlying fault conditions are resolved.
This is particularly crucial in production, where many failures are transient~\cite{DBLP:conf/hotos/HuangGZLDCY17,DBLP:conf/osdi/HuangGLZD18,DBLP:conf/nsdi/Chen0NYX23}.
Issues like temporary network congestion or resource contention can cause service slowdowns or intermittent errors that may resolve on their own shortly. A resilient system must be able to withstand this temporary degradation without causing a complete outage.
Moreover, the interconnected nature of microservice systems means that failures in individual components can propagate through the dependency chains, potentially affecting multiple services.
In this case, when the faulty component is repaired, the affected services must be able to autonomously return to a healthy operational state.


Despite its importance, effective microservice resilience testing presents significant challenges. Traditional methods are limited in two key aspects.
First, they often require extensive manual effort~\cite{DBLP:conf/icdcs/HeorhiadiRJRS16,DBLP:journals/jss/WaseemLSSM21,DBLP:conf/issta/YangLS0FYL24}, rendering them impractical for large-scale systems. This includes manually designing test cases, configuring complex fault injection scenarios, and defining appropriate assertions beyond simple status checks. Such a manual process is slow, error-prone, and cannot scale with the complexity and rapid evolution of modern services.
Second, their applicability to industrial systems is limited by the insufficient expressiveness of both fault models and verification mechanisms.
Many studies~\cite{DBLP:conf/qrs/GiamatteiGPR22,DBLP:journals/tse/WuYNNPHY23,DBLP:journals/tdsc/ChenCYLH24} rely on small benchmarks (e.g., TrainTicket~\cite{DBLP:journals/tse/ZhouPXSJLD21,trainticket}, SockShop~\cite{sockshop}) that lack the scale and intricate dependencies of production environments.
The injected faults are often generic (e.g., basic HTTP errors) and validated with simple checks (e.g., status code 200), providing limited ability for testers to create customized checks. Such approaches fail to capture the rich spectrum of real-world failures like specific middleware exceptions or complex asynchronous issues~\cite{DBLP:conf/icws/LongWCCCW20,DBLP:conf/ast/CamilliGJRR22,DBLP:journals/tpds/WangGGBY18}.

To address these limitations, we present \alias, an end-to-end framework for scalable and effective microservice resilience testing in industrial environments. Our approach has three key design goals.
First, for \textbf{production-level fidelity}, \alias combines online traffic record-and-replay with mocking technique for fine-grained, application-level fault injection.
To ensure that recorded traffic can be successfully replayed~\cite{DBLP:conf/icse/LiuLDLZ22}, \alias dynamically identifies and updates state-dependent variables to reflect the current execution context.
Moreover, \alias constructs a fault library of realistic failure modes (including software-level exceptions and communication errors) to rigorously exercise a service's internal error-handling logic.
Second, for \textbf{scalability}, \alias tackles the combinatorial explosion of the test space through a complexity-driven strategy.
It intelligently selects a small yet potent set of test cases that target the most intricate and failure-prone execution paths.
This provides a systematic solution to proactively discover critical resilience vulnerabilities.
Finally, for \textbf{automation}, \alias orchestrates the entire test execution lifecycle through a three-phase pipeline (i.e., startup, fault injection, and recovery), while automatically verifying the system's resilience against multi-faceted criteria.

We have deployed \alias in Huawei Cloud for over eight months, serving a large number of microservice applications for their automated resilience testing.
These applications span various business domains including storage, networking, digital platform, and Internet of Things (IoT), etc.
Through our long-term deployment, we have gained valuable insights into the practical challenges and trade-offs of resilience testing at scale.
In this paper, we report results for four large-scale and representative applications, which comprise hundreds of microservices.
They are characterized by high-volume traffic and complex architectures with intricate service dependencies and a mix of technologies.
Overall, \alias uncovers 137 potential vulnerabilities, out of which 89 have been confirmed through developers' manual investigation at the time of writing.
Furthermore, in a controlled experiment against a benchmark of 48 reproduced bugs, \alias achieves a 90\% detection coverage.
These results demonstrate the effectiveness of our framework in improving the reliability of microservice systems in industrial environments.

In summary, we make the following major contributions:

\begin{itemize}[noitemsep,leftmargin=5.5mm]
    \item We present \alias, an automated, end-to-end framework for microservice resilience testing that combines replay-based fault injection with a complexity-driven strategy to prioritize failure-prone execution paths and prune redundant tests. \alias orchestrates the entire testing lifecycle and employs a multi-faceted oracle to automatically assess resilience against both final outcomes and internal health states.


    
    \item We demonstrate the practical effectiveness of \alias through a large-scale deployment in Huawei Cloud over eight months, where it has been used by many development teams. We report on its long-term effectiveness and experimental results on reproduced vulnerabilities, validating its real-world impact.

    \item We share critical lessons learned from industrial deployment about the importance of deep verification oracles that go beyond API boundaries, the trade-offs between perfect test prioritization and automation efficiency, and practical considerations for integration into existing development workflows.
\end{itemize}

\section{Background}
\label{sec:background}

\subsection{Microservice Systems}

The microservice architectural style has become the de facto standard for building large-scale cloud applications~\cite{fowler2014microservices}. Unlike monolithic applications, a microservice system is composed of a collection of small, autonomous services, each responsible for a specific business capability. These services are independently developed, scaled, and maintained, offering organizations greater agility and flexibility. They typically communicate with each other over a network using lightweight protocols such as HTTP/REST or gRPC. The overall behavior of the system emerges from the intricate interactions between these services, making it difficult to reason about and predict. A fault in a single service can trigger a cascade of failures that propagates throughout the system, leading to widespread outages.

Microservices do not operate in isolation. They rely heavily on a shared backbone of platform components to manage state, communication, and data persistence. This includes \textit{databases} (e.g., MySQL, MongoDB) for persistent storage, \textit{message brokers} (e.g., Kafka, Pulsar) for asynchronous communication and event-driven workflows, and \textit{in-memory caches} (e.g., Redis, Memcached) for performance and latency.
While these components are engineered for high availability, their complex interactions with microservices introduce significant risks for error propagation.
The application's ability to handle components unavailability, connection errors, or data serialization issues is paramount to its overall resilience. Thus, a modern microservice system forms an interdependent network of business services and infrastructure, where resilience depends on the robustness of every component and every interaction.

\begin{figure*}[t]
    \centering
    \includegraphics[width=0.87\linewidth]{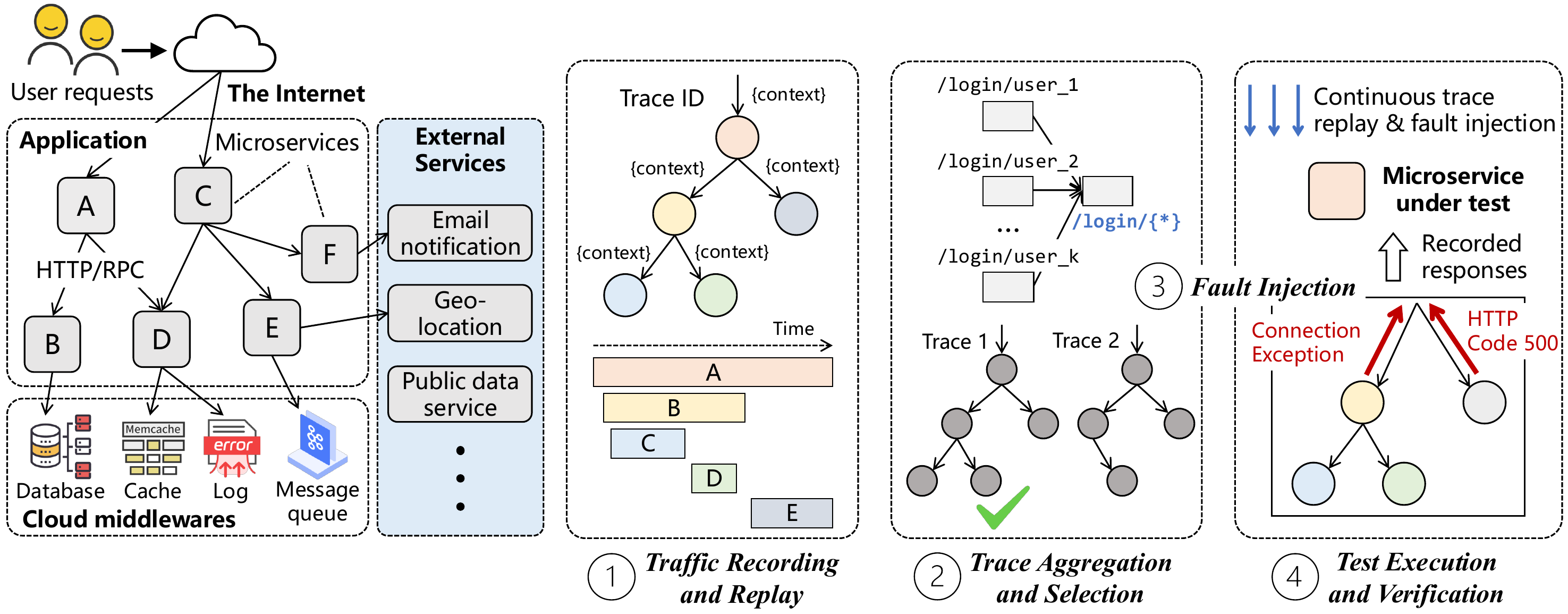}
    \caption{Overall Framework of \alias for Automated Microservice Resilience Testing}
    \label{fig:framework}
\end{figure*}

\subsection{Resilience Testing of Microservice Systems}
\label{sec:Resilience Testing of Microservice Systems}

Resilience is the ability of a system to withstand failures and maintain an acceptable level of service. For microservice systems, this means gracefully handling the inevitable failures of services, network, and platform components.
Resilience testing (popularized under the term \textit{Chaos Engineering}) aims to proactively discover vulnerabilities by injecting faults to build confidence in a system's survival under volatile production conditions.
Existing approaches can be broadly categorized into two streams based on the granularity of fault injection.
\textit{Infrastructure-level approaches}~\cite{DBLP:journals/tse/WuYNNPHY23} focus on manipulating the environment, such as terminating VMs, introducing latency, or dropping packets. While these faults effectively exercise basic resilience tactics like retries and circuit breakers for connectivity issues, they are often too coarse-grained to trigger the application-specific error-handling logic tied to internal service states.
In contrast, \textit{application-level approaches} target business logic to verify service fallbacks or recovery procedures. For instance, simulating specific business-level exceptions (e.g., partial API failures or platform errors) requires this finer granularity.

To achieve high fidelity in application-level testing, researchers and practitioners often use two common and often complementary techniques, i.e., traffic record-and-replay and mocking. Record-and-replay generates realistic workloads by capturing and replaying production traffic, ensuring test scenarios reflect actual user interactions. Mocking provides fine-grained control by replacing downstream dependencies with simulated versions that can return programmed failures.
However, approaches~\cite{DBLP:conf/sigsoft/LongWC0020,DBLP:conf/kbse/Arora0IKR18,DBLP:conf/msr/SpadiniABB17,DBLP:journals/ese/SpadiniABB19,DBLP:conf/sigsoft/ArcuriFG15} leveraging these techniques often face the following critical challenges:

\begin{itemize}[noitemsep,leftmargin=5.5mm]
    \item \textbf{State Dependency}: Production traffic is stateful and time-sensitive. Directly replaying recorded requests may fail due to stale data, such as expired timestamps, single-use idempotency keys, or invalid tokens.
    An effective replay system must intelligently identify and refresh these dynamic variables to reflect the current testing context, ensuring the requests are valid.

    \item \textbf{Scalability}: The immense volume of production traffic makes replaying every request infeasible.
    Furthermore, the number of potential fault injection points (e.g., every service-to-service call, every database query) creates a combinatorial explosion of possible test cases. Naive or random selection of tests is inefficient and may miss critical, yet infrequent, execution paths.

    \item \textbf{Automation}: Fully automating the resilience testing lifecycle presents a significant operational challenge. It requires the complex orchestration of setting up test environments, initiating workloads, and synchronizing fault injections. Another core difficulty is the test oracle problem, i.e., how to automatically and reliably determine whether a test has passed or failed.
\end{itemize}

\section{Methodology}
\label{sec:methodology}

\subsection{Overview}

To address these challenges, we present \alias, an automated framework for effective microservice resilience testing, targeting Java-based applications.
The overall framework of \alias is shown in Fig.~\ref{fig:framework}, which consists of four phases, i.e., \textit{traffic recording and replay}, \textit{trace aggregation and selection}, \textit{fault injection}, and \textit{test execution and verification}.
In the first phase, \alias captures live production traffic and applies a lightweight heuristic to identify and parameterize state-dependent variables, transforming raw traces into replayable test templates. To address the immense volume of traffic, the second phase employs a complexity-driven strategy, aggregating similar user interactions and selecting a small set of test cases that target the most failure-prone execution paths. The third phase includes the design of a comprehensive library of realistic, application-level faults and a principled pruning mechanism to identify the most impactful injection targets. The final phase automates the entire test run through a three-stage pipeline and a multi-faceted verification oracle for assessing the system's resilience.



\subsection{Traffic Recording and Replay}
\label{sec:traffic_recording}

\alias begins by recording production traffic, which is reconstructed as distributed traces composed of causally-related spans.
While industry standards like OpenTelemetry~\cite{opentelemetry} provide capabilities for distributed tracing, they are primarily designed for passive monitoring.
\alias requires not only execution recording but also active runtime intervention to support traffic replay and fault injection (e.g., intercepting a database call to return a mock response or throw an exception).
Thus, we implement a non-intrusive instrumentation technique using a dynamic AOP framework based on Java agents~\cite{DBLP:conf/icse/LiuLDLZ22,DBLP:conf/asplos/LiuLLLLGT17,jvm-sandbox}.
This allows us to intercept method calls at runtime without modifying source code, capturing comprehensive operations including inter-service communications (HTTP/RPC) and interactions with platform components, such as databases (via JDBC), message brokers (e.g., Kafka, Pulsar), and in-memory caches (e.g., Redis). 
These traces serve as the fundamental artifacts for our subsequent analysis, high-fidelity replay, and resilience testing.

However, as discussed in Sec.~\ref{sec:Resilience Testing of Microservice Systems}, directly replaying traces often fails due to time-sensitive and state-dependent variables.
For example, a request containing an outdated timestamp may get rejected.
Also, many API calls include an idempotency key to prevent duplicate processing.
Replaying a request with a stale key would be ignored.
Liu et al.~\cite{DBLP:conf/icse/LiuLDLZ22} studied this problem by categorizing these variables as system states, internal states, and external states, and proposed an approach to identify them based on static taint analysis.
However, such an approach is not scalable for large service codebases and is tightly coupled to the specific frameworks adopted in~\cite{DBLP:conf/icse/LiuLDLZ22}, i.e., SOFA/SOFABoot.
To address this, we introduce a lightweight dynamic variable identification technique that operates directly on the recorded request and response payloads.
The core of our approach is a two-stage heuristic designed to distinguish dynamic variables from static data without analyzing source code:

\begin{itemize}[noitemsep,leftmargin=5.5mm]
    \item \textbf{Intra-span Correlation}: For each recorded span, we first analyze its request and response payloads to find tokens that appear verbatim in both. The intuition is that many state-dependent variables, such as session IDs or transaction tokens, are often received in a request and propagated directly into the response to maintain context.
    \item \textbf{Inter-span Variability}: To filter out static values (e.g., a hardcoded version string), we then compare these candidate tokens across multiple spans of the same operation. A token is confirmed as a dynamic variable only if its value differs across these independent instances.
\end{itemize}

\begin{figure}[t]
    \centering
    \includegraphics[width=0.94\linewidth]{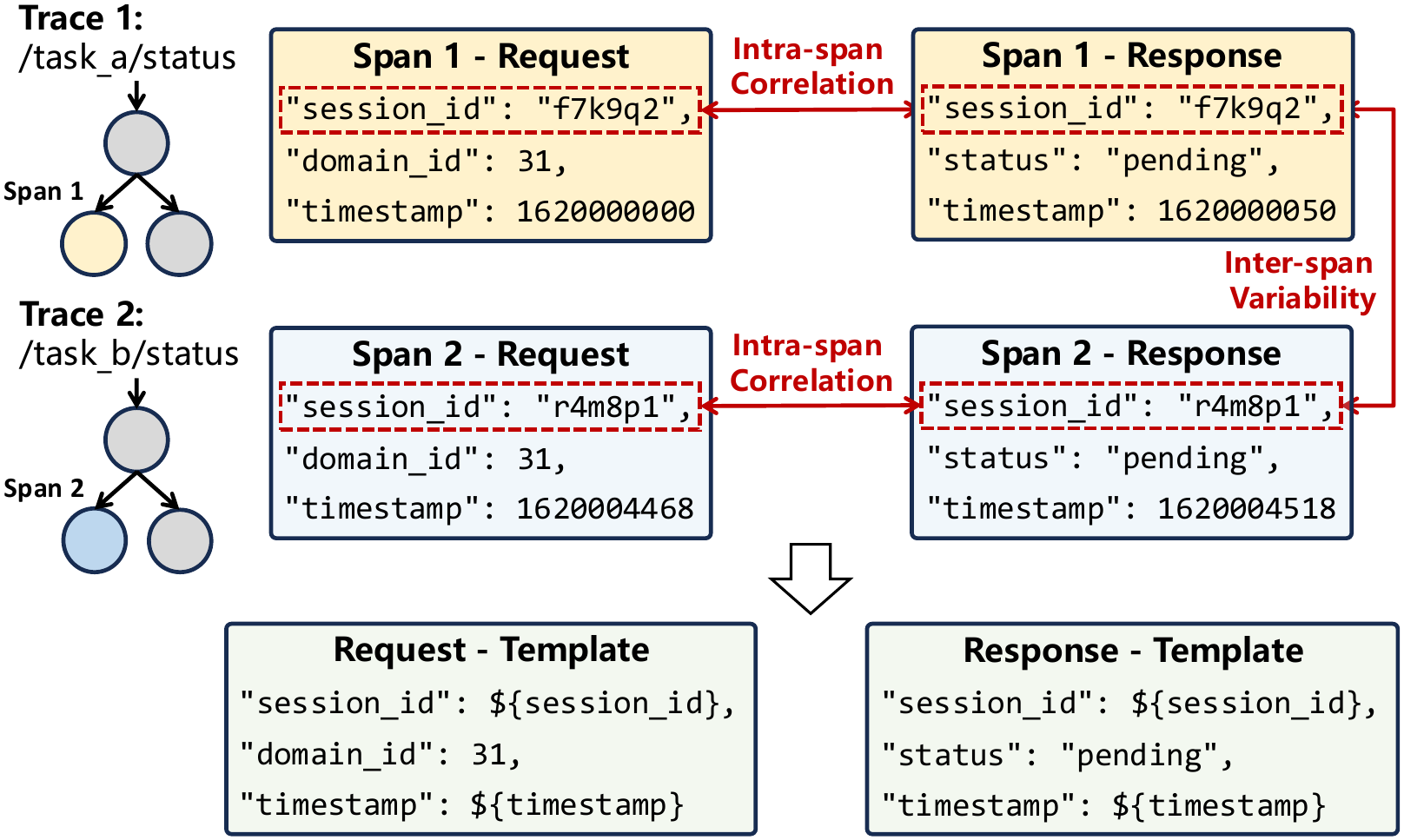}
    \caption{Trace Templating in Traffic Replay}
    \label{fig:trace_templating}
\end{figure}

Fig.~\ref{fig:trace_templating} provides a visual walkthrough using two distinct spans recorded from the same logical operation. In the first stage, \textit{Intra-span Correlation}, we identify candidate tokens like \texttt{session\_id} that appear in both the request and response of a single span. In the second stage, \textit{Inter-span Variability}, we compare these candidates across independent spans. We observe that \texttt{session\_id} changes (e.g., \texttt{"f7k9q2"} vs. \texttt{"r4m8p1"}), confirming it is dynamic. Conversely, fields like \texttt{domain\_id} and \texttt{status} remain constant across both spans and are therefore treated as static data.
Tokens that satisfy both conditions are abstracted into parameterized placeholders (e.g., \texttt{\$\{*\}}), transforming recorded traffic into a reusable traffic template.
During replay, these placeholders are automatically instantiated based on the test's execution context: unique identifiers like \texttt{session id} receive newly generated values, and timestamps are substituted with the current system time.
This payload-centric approach treats services as black boxes, making it fundamentally more scalable and framework-agnostic than static taint analysis.

\subsection{Trace Aggregation and Selection}
\label{sec:trace_aggregation}

After establishing replayable traffic, 
the next step is to select a representative set of traces for resilience testing, which aims to address two scalability challenges.
The first is the immense volume of traces from production-grade applications, e.g., millions of requests, rendering exhaustive testing impractical.
This volume largely stems from redundancy, where repetitive calls to the same interface differ only in parameterized values.
For instance, thousands of distinct traces might be generated when different users log in, all invoking the same \texttt{POST /api/login/\{username\}} API.
Thus, to avoid redundant testing, we perform traffic aggregation to partition the entire corpus of traces into distinct categories based on their entry point interfaces.
Specifically, we employ Drain~\cite{DBLP:conf/icws/HeZZL17}, an algorithm commonly used in log parsing, to dynamically learn the static and variable parts of the root span's request line, which typically includes the HTTP method and URI path. This step effectively distills the vast number of raw calls into a set of distinct service interfaces.

Even after traffic aggregation, we are still facing another scalability problem, i.e., production applications often contain tens or even hundreds of microservices, each of which exposes multiple interfaces.
Exhaustively replaying each unique trace to test all interfaces remains infeasible.
Instead of random sampling, we develop a heuristic-based scoring mechanism to select a subset of traces that are more likely to reveal resilience issues.
While sophisticated algorithms like lineage-driven fault injection (LDFI)~\cite{DBLP:conf/sigmod/AlvaroRH15} exist, they require detailed system models and static analysis that are impractical in our multi-team, polyglot environment with hundreds of rapidly-evolving services.
Our approach starts by quantifying \textit{trace complexity} using a weighted combination of three factors: trace length measured by the number of spans, the diversity of components (i.e., the unique services and infrastructure components involved), and the end-to-end duration of the trace execution.
The weights are configurable based on specific testing priorities.
Traces involving more diverse components and longer execution paths receive higher complexity scores, as they exercise more interactions and are inherently more failure-prone.
This per-trace complexity score serves as a building block for our next level of selection by prioritizing microservice interfaces.
An interface's complexity can vary significantly based on input parameters and business logic, leading to different execution paths.
To capture a holistic measure of an interface's typical complexity, we calculate an aggregate score for it by averaging the complexity scores of all associated traces.

With each interface now assigned a complexity score, we employ a two-level selection strategy to identify the final set of test cases. First, we rank all interfaces by their aggregate score and select the top-K candidates.
Second, for each candidate, we select its single most representative test case, i.e., the trace with the highest individual complexity score.
This strategy concentrates testing efforts on the system's most complex interfaces and their most challenging execution paths, resulting in a small yet highly potent set of traces for the subsequent fault injection phase.



\subsection{Endpoint Coverage-guided Fault Injection}
\label{sec:phase3-fault_injection}

\subsubsection{Fault Library Construction}

The foundation of our fault injection is a comprehensive and extensible fault library that catalogs realistic failure modes. 
In particular, we focus on software-level faults to test an application's internal error-handling logic, instead of coarse-grained infrastructure failures.
Using dynamic instrumentation, we create injectable fault modules that simulate the following two primary categories of failures:

\begin{itemize}[noitemsep,leftmargin=5.5mm]
    \item \textit{Platform components exceptions}: This category simulates common exceptions thrown by client libraries of different platform components.
    To build a comprehensive list, we conduct a systematic analysis that combines: 1) a thorough review of official API documentation to identify documented failure modes; 2) an empirical analysis of historical incidents to prioritize high-impact errors; and 3) interviews with senior developers and SREs to capture expert domain knowledge. This has yielded a rich library of faults targeting various platform component categories, such as database (e.g., \texttt{SQLTimeoutException}), message brokers (e.g., \texttt{SerializationException}), and cache (e.g., \texttt{RedisConnectionException}). This fault library is not static but is continuously updated to reflect changes in our technology stack and to incorporate newly identified failure patterns.

    \item \textit{Inter-service communication errors}:
    This category models synchronous communication (e.g., HTTP/RPC) failures by intercepting outgoing requests to simulate a wide range of disruptions. This includes introducing network delay (latency) to test for slow dependencies or throwing protocol-level exceptions like \texttt{java.net.SocketTimeoutException} to simulate a complete connection failure. Moreover, instead of making the actual network call, they can return a fully manipulated failure response. Our library provides extensive coverage for this, allowing for the alteration of the entire response, including standard HTTP status codes, such as Client Errors (e.g., \texttt{401 Unauthorized}) and Server Errors (e.g., \texttt{500 Internal Server Error}, \texttt{504 Gateway Timeout}), and the response body itself.
\end{itemize}

\subsubsection{Fault Injection Target Identification}

Based on the selected traces (Sec.~\ref{sec:trace_aggregation}) and fault library, the next step is to determine where to inject the faults, i.e., \textit{fault injection targets}.
To achieve comprehensive fault coverage, we introduce a fine-grained abstraction for injection targets called an \textit{endpoint}.
An endpoint is a tuple (\texttt{Component}, \texttt{Framework}, \texttt{Method}): \texttt{Component} defines the semantic context of the interaction (e.g., Database, Cache);
\texttt{Framework} identifies the specific library or implementation (e.g., MyBatis, Jedis), and \texttt{Method} specifies the operation (e.g., update, get).
This abstraction allows us to distinguish functionally identical operations implemented via different frameworks in different teams, each possessing unique failure-handling characteristics.
For instance, we can differentiate critical endpoint pairs such as \texttt{RPC-gRPC-findUserById} vs. \texttt{RPC-Dubbo-findUserById} for user lookups, or \texttt{MQ-Kafka-send} vs. \texttt{MQ-Pulsar-send} for messaging operations.
This level of precision is essential for systematically discovering implementation-specific vulnerabilities.
Each endpoint serves as a target for relevant faults from our library, allowing testing various failures for a single operation. For example, an \texttt{MQ-Kafka-send} endpoint can be injected with a \texttt{TimeoutException} to test for network issues or with a \texttt{SerializationException} to simulate a data format mismatch.

Technically, every span of a trace corresponds to one endpoint, as it records an operation performed within the microservice system (Sec.~\ref{sec:traffic_recording}).
However, as a single complex trace can contain hundreds or even thousands of spans, injecting faults into all its endpoints creates a combinatorial explosion of potential test cases.
To address this, \alias employs a principled pruning mechanism to systematically reduce the test space by eliminating redundant tests and prioritizing high-value tests through the following three strategies:

\begin{figure}[t]
    \centering
    \begin{subfigure}[b]{\linewidth}
        \centering
        \includegraphics[width=0.9\linewidth]{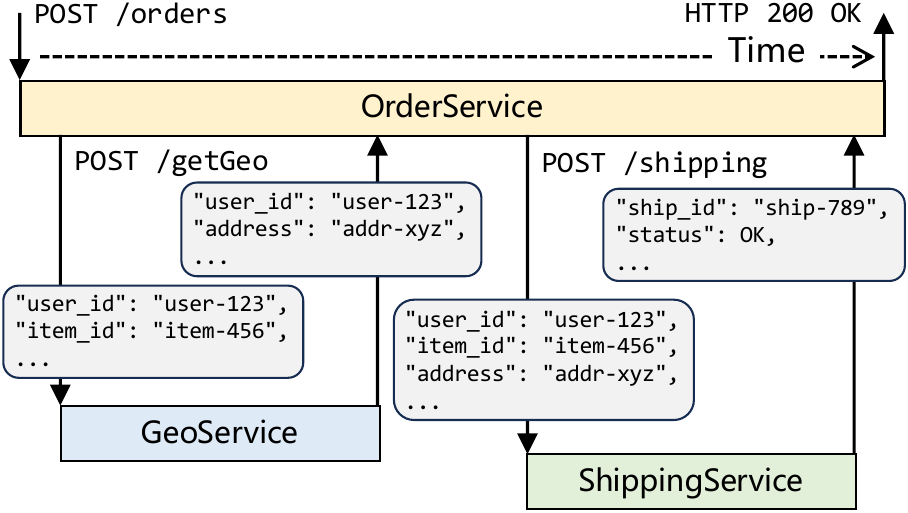}
        \caption{Producer-consumer Data Dependency Pattern}
        \label{fig:consumer-producer}
    \end{subfigure}
    
    \vspace{0.4cm}
    
    \begin{subfigure}[b]{\linewidth}
        \centering
        \includegraphics[width=0.9\linewidth]{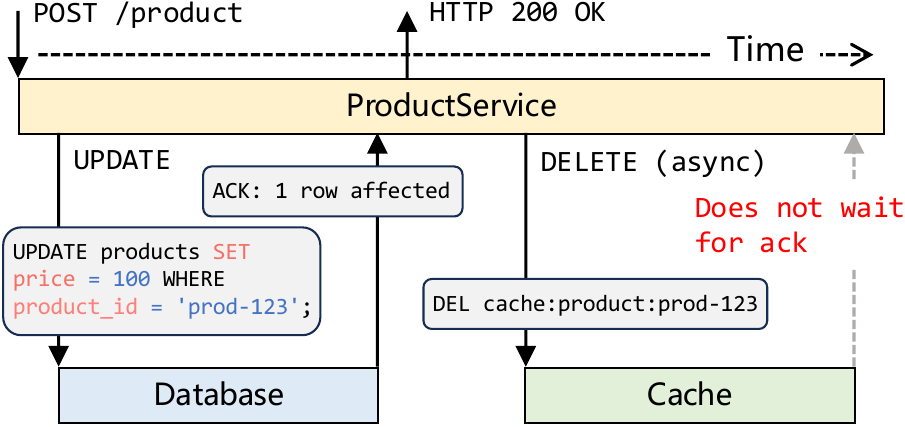}
        \caption{Dual-write Data Dependency Pattern}
        \label{fig:dual-write}
    \end{subfigure}

    \caption{Data-flow Dependency Patterns}
    \label{fig:dataflow_dependency}
\end{figure}

\begin{figure*}[t]
    \centering
    \includegraphics[width=0.86\linewidth]{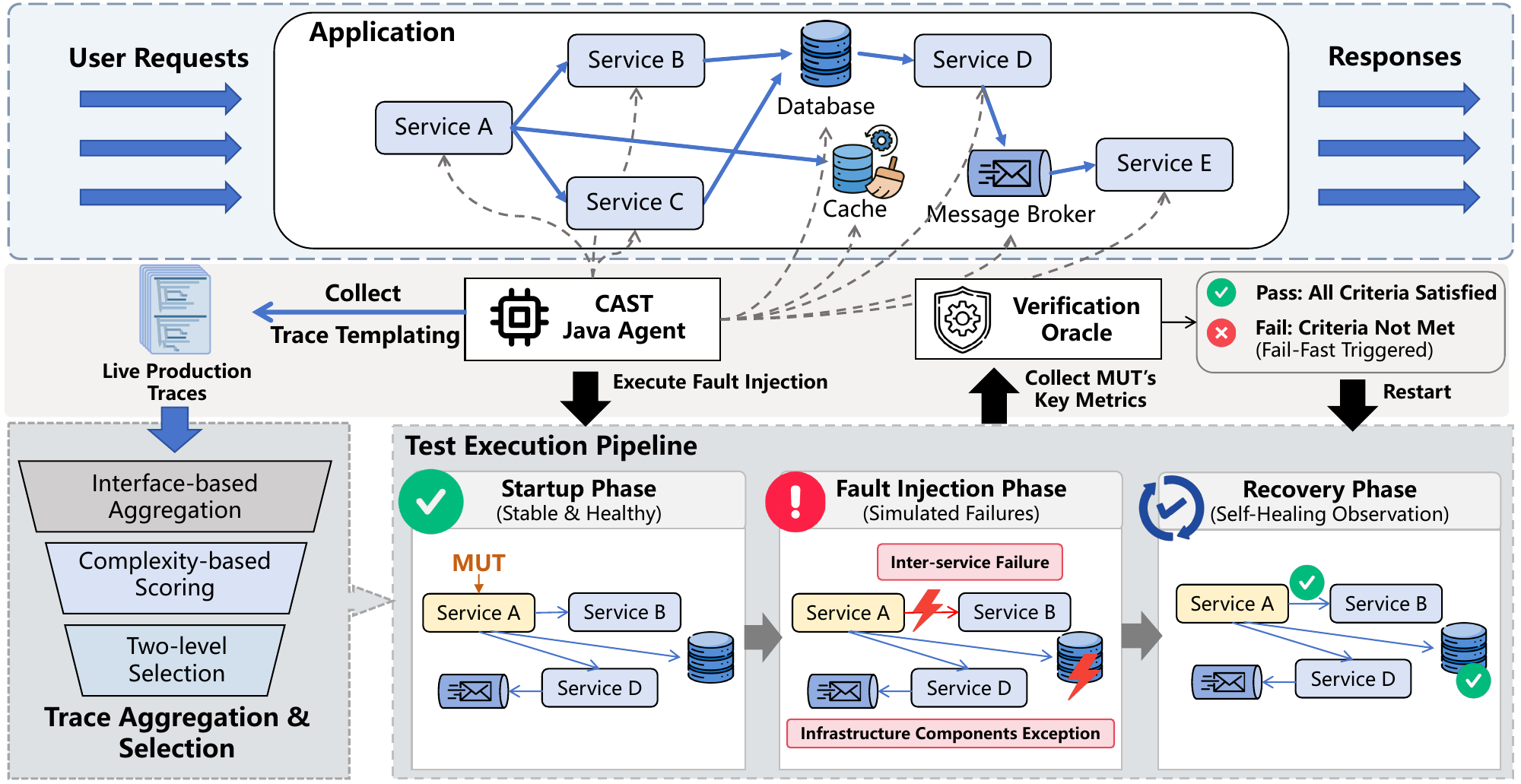}
    \caption{Test Execution Pipeline of the \alias System}
    \label{fig:cast_runtime}
\end{figure*}

\begin{itemize}[noitemsep,leftmargin=5.5mm]
    \item \textit{Cross-service sampling}: The fault-handling logic for a given endpoint (e.g., a database call) is often implemented in a shared library or framework, making its behavior consistent across different services. It is inefficient to exhaustively test this same endpoint in every microservice that invokes it.
    Therefore, we randomly sample a small, configurable number of distinct microservices (e.g., three) for fault injection.
    This approach recognizes that resilience issues may still manifest differently across implementations despite using identical frameworks.
    
    \item \textit{Intra-trace redundancy elimination}: A single trace may contain multiple spans of the same endpoint (e.g., repeated database queries). To avoid redundancy, 
    we inject faults only at the last invocation. The rationale is that the final invocation operates on a more advanced system state, influenced by all preceding operations, making it more likely to expose complex, state-dependent vulnerabilities.
    Furthermore, for database transactions, faults at the last invocation test the system's ability to rollback a fully-staged transaction, which is typically more error-prone than rolling back earlier, partial operations.

    \item \textit{Data-flow dependency prioritization}: 
    This strategy prioritizes injection targets by analyzing the structure of data-flow dependencies within traces, as they are prone to causing cascading failures.
    Based on the data overlap in payloads and operation sequences, we identify two fundamental dependency structures, as shown in Fig.~\ref{fig:dataflow_dependency}.
    The first pattern, \textit{Producer-Consumer} (Fig.~\ref{fig:consumer-producer}), 
    involves a service (e.g., OrderService) first calling a component to retrieve data (e.g., calling GeoService to validate an address) 
    and then using that produced data in a subsequent call to another component (e.g., calling ShippingService with the validated address). 
    The integrity of this entire workflow depends on the successful flow of data from the producer (GeoService) to the consumer (ShippingService). 
    We prioritize the producer as a high-value injection target because its failure can propagate through the data chain, allowing us to skip testing downstream consumers in isolation for the same trace.
    The second pattern, \textit{Dual-Write} (Fig.~\ref{fig:dual-write}), occurs when the same set of data in a service (e.g.,  the product ID and price in ProductService) is used to invoke two or more separate operations (e.g., \texttt{UPDATE} in the Database and \texttt{DELETE} in the Cache).
    This pattern is common during data synchronization tasks and is highly susceptible to silent but critical inconsistency bugs.
    Particularly, the second write to the cache is often performed asynchronously for performance reasons, where the service does not wait for an ack.
    In this case, we bypass the primary synchronous write and focus exclusively on the secondary/asynchronous operation. 
    Since the primary write is typically covered by the Producer-Consumer or individual service tests, focusing on the secondary write targets the specific source of silent data inconsistency while significantly reducing the number of required test cases.
    As discussed in Sec.~\ref{sec:phase4_test_oracle}, this scenario necessitates a more sophisticated verification oracle.
\end{itemize}

This principled pruning mechanism yields a set of fault injection test cases with strategic diversity, covering a wide spectrum of operations and frameworks in microservices.
It allows us to validate the common fault-handling patterns without the cost of exhaustive testing, striking a balance between coverage and efficiency.

\subsection{Automated Test Execution and Verification}
\label{sec:phase4_test_oracle}

With the fault injection test cases, the final phase involves their systematic execution and the verification of the system's resilience.

\subsubsection{Execution Pipeline and Test Scheduling}

Each individual test case is executed within a well-defined, three-phase cycle. 
Fig.~\ref{fig:cast_runtime} provides a runtime perspective of this process, detailing how the Java agent interacts with the microservice under test (MUT) and platform components during each phase.
Particularly, each phase is configured with a user-defined duration and injection frequency to ensure sufficient time for the system to stabilize and for effects to be observed.
In this process, we continuously generate traffic against the MUT by replaying the selected trace.

\begin{itemize}[noitemsep,topsep=0pt,leftmargin=5.5mm]
    \item \textit{Startup phase}: 
    The MUT is deployed as a standalone, on-demand container. To ensure strict data isolation and avoid polluting shared environments, we initialize a fresh instance for each test. No faults are injected in this phase to ensure that the system is operating in a stable, healthy state.
    \item \textit{Fault injection phase}: Selected faults are activated, targeting the designated endpoints. This allows us to monitor the system's behavior under specific failure conditions.
    \item \textit{Recovery phase}: Fault injection is deactivated. This final phase is crucial for observing whether the system can autonomously recover and return to its original healthy operational state.
\end{itemize}

A practical challenge in this process is the overhead associated with the MUT's startup. 
This process can be time-consuming as it involves the initialization of the application and environments, as well as the time for the service to reach its stable operational state. Executing only one fault injection per startup would be prohibitively slow.
To amortize this cost, \alias groups multiple fault injection tests into a single run.
We employ a greedy scheduling algorithm that iteratively selects traces providing the maximum increase in new endpoint coverage. This minimizes the number of required startups to execute all fault injection tests.
Within a single test run, these batched tests are executed sequentially.

However, such a design introduces a potential problem of state contamination, where a fault from a preceding test alters the system state and influence subsequent tests. To mitigate this, \alias adopts a ``fail-fast'' execution model. Specifically, as long as each test passes, the pipeline proceeds to the next one. When a test fails, indicating a potential vulnerability, the entire run is halted immediately. The system is then fully restarted for the next test, ensuring that the remaining tests start from a clean state.
We acknowledge the possibility that even a previous passing test could subtly alter system state. However, we consider this a pragmatic trade-off for the significant efficiency gains. In practice, such scenarios are rare, and our cross-service sampling strategy provides inherent redundancy.
Furthermore, to enhance long-term effectiveness across multiple testing cycles, \alias maintains a history of executed test cases. When scheduling new tests from subsequent batches of recorded traffic, this history is used to avoid re-executing any identical tests.

\subsubsection{Verification Oracle}

To automatically determine the outcome of a test, \alias employs a multi-faceted verification oracle with the following two key components:

\begin{itemize}[noitemsep,leftmargin=5.5mm]
    \item \textit{Phase-based performance criteria}: Our oracle defines a set of performance criteria, which are quantifiable assumptions about the MUT's key metrics (e.g., success rate, latency, and throughput). Specific criteria values are derived from historical trace data, which can vary depending on the execution phase. For instance, the startup phase requires a 100\% success rate to confirm correct system initialization. To verify that the fault is having an impact, the success rate during the fault injection phase is expected to drop significantly (e.g., $\le$ 30\%). During the recovery phase, the success rate must return to a healthy level (e.g., $\ge$ 80\%) to demonstrate the system's ability to self-heal. This threshold is intentionally set below 100\% to account for a realistic recovery time window as the system stabilizes.
    While this trade-off between automation and absolute precision may occasionally cause false positives or negatives (discussed in Sec.~\ref{sec:rq2-effectiveness}), it enables practical automated verification.

    \item \textit{Granular assertion points}:
    A naive approach of checking only the final HTTP response code at the service entry point is often insufficient. We observed that for certain asynchronous operations, such as publishing a message to Pulsar, the call might return successfully to the application logic even if the Pulsar subsequently fails to deliver the message to the consumer. This ``fire-and-forget'' behavior, where the service considers the task complete once submitted without awaiting its actual delivery, would lead to an incorrect pass judgment. would lead to an incorrect pass judgment. To solve this, \alias establishes assertion points not only at the service entry point but also directly at the internal endpoint where the fault is injected. This dual-level verification ensures we accurately assess both the fault's direct impact and its effect on the service's overall behavior.
    
\end{itemize}

A test is considered passed only when the observed behavior meets the defined criteria for all three phases across all relevant assertion points. Failed tests are flagged for further investigation.


\section{Evaluation}
\label{sec:evaluation}

In this section, we evaluate the performance of \alias in our production deployments. In particular, we aim to answer the following two research questions:

\begin{itemize}[noitemsep,leftmargin=5.5mm]
    
    \item \textbf{RQ1 (Replayability)}: How effective is \alias in making recorded traffic replayable?

    \item \textbf{RQ2 (Effectiveness)}: How effective is \alias in automatically discovering resilience vulnerabilities?
\end{itemize}

\alias is deployed as a centralized resilience testing platform within Huawei Cloud. For over eight months, \alias has been used by many development teams to proactively identify resilience issues in their applications.
Given \alias's role as a shared platform where teams can independently run tests and analyze results, it is impractical to manually track and confirm the outcome of every potential vulnerability reported across all service teams. Therefore, for this paper, we focus our analysis on the testing results from four representative applications (anonymized as Service 1-4). They are specifically chosen because their characteristics align with the challenges \alias is designed to address, i.e., large scale (hundreds of microservices), high volume of business-critical traffic, and architectural complexity with intricate dependencies.

\subsection{RQ1: The Replayability of Recorded Traffic}

To answer RQ1, we evaluate the effectiveness of \alias's dynamic variable identification technique in making production traffic replayable. A successful replay is defined as a replayed request that is accepted by the target service and returns an expected success code (e.g., HTTP 2xx), establishing a baseline before fault injection.

We begin by collecting six hours of traffic (over three million traces) from the target applications.
\alias's aggregation module distills this corpus into 291 unique service interface calls. For each interface, \alias analyzes all of its associated trace instances to apply the two-stage heuristic (Sec.~\ref{sec:traffic_recording}), automatically identifying and parameterizing the time-sensitive and state-dependent variables. To validate the outcome, we randomly select one trace instance for each of the 291 interface calls and attempt to replay it against the live test system.
Out of the 291 replay attempts, \alias successfully replays 286 traces, improving the success rate from below 80\% (for naive replay) to 98.3\% with our templating technique.
This high success rate demonstrates that our automated approach is highly effective at handling the state-dependency problem for the vast majority of interfaces.
Repeated experiments with different random samples yield consistent results, confirming the method's stability.


We manually investigate the 5 failed cases and find that they are primarily caused by highly business-specific or system-specific variables that escape our general-purpose identification method. For instance, some requests include dynamically computed security signatures or tokens derived from complex, proprietary business logic not reflected in the request-response payload structure. To address these edge cases in a practical industrial setting, \alias maintains a configurable list, allowing engineers to manually register or deregister variables for a given service. This hybrid approach ensures that comprehensive replay coverage can be achieved with minimal, one-time manual effort for these outliers.






\subsection{RQ2: The Effectiveness of Resilience Bug Detection}
\label{sec:rq2-effectiveness}

To evaluate \alias's effectiveness, we first consider its long-term impact in real-world production systems. Over an eight-month deployment in Huawei Cloud, \alias has been continuously adopted by different teams to improve service reliability.
For this paper, we present the results of \alias's utilization across four large-scale and representative applications.
Overall, \alias successfully uncovers 137 potential vulnerabilities, of which 89 have been confirmed through developers' manual investigation at the time of writing. This demonstrates \alias's practical value in production.

\begin{figure}[t]
    \centering
    \includegraphics[width=0.92\linewidth]{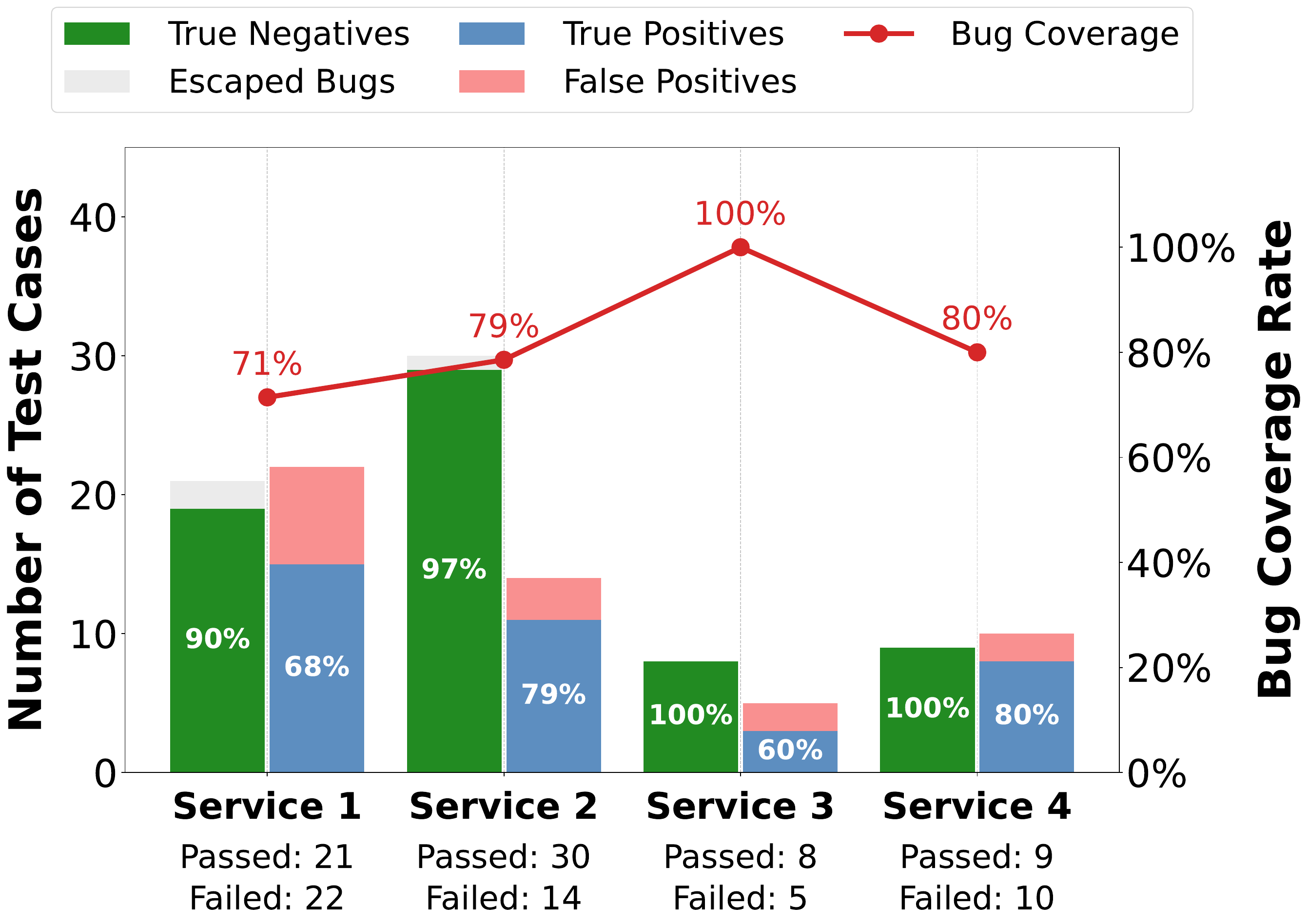}
    \vspace{-2pt}
    \caption{Experimental Results on Different Services}
    \label{fig:benchmark_results}
    \vspace{-6pt}
\end{figure}
 
For a more controlled, quantitative analysis of its detection capability, we collect a ground truth dataset of 48 confirmed vulnerabilities. These cases consist of historical defects identified through incident reports and manual investigation.
It provides a comprehensive benchmark for our experiment. 
We use the same six-hour corpus of recorded traffic from RQ1 to generate test cases against the systems containing these bugs. Following our complexity-driven selection strategy (Sec.~\ref{sec:trace_aggregation}), we select the top-150 interfaces with the highest complexity scores and their representative traces. We then apply our pruning logic (Sec.~\ref{sec:phase3-fault_injection}) to identify endpoints.
For each type of endpoint, we randomly select at most three different microservices for testing.
This process yields 119 distinct test cases that cover 477 fault injection targets.

The detailed results in Fig.~\ref{fig:benchmark_results} show that upon executing the 119 test cases, 68 passed and 51 failed.
\alias demonstrates a high true negative rate, meaning nearly all passed cases are indeed free of bugs.
Manual investigation of the 51 failed tests reveals that \alias successfully identifies 37 of the 48 known vulnerabilities, achieving a consistently high bug coverage across different services (77.1\% in aggregate).
The remaining 14 failed tests are identified as false positives. In these cases, faults are correctly injected and the system's behavior changes, but this does not represent a genuine resilience bug.
Detailed analysis reveals that 9 of these false positives are caused by our default performance thresholds being overly sensitive for certain non-critical background operations (e.g., metrics collection, logging), while 5 are due to legitimate degraded-but-acceptable behavior under fault conditions.
Similarly, for the 3 false negatives, a test case is executed but the oracle incorrectly reports a ``pass.''
This typically occurs when the impact of a fault is too subtle to violate the default performance thresholds, e.g., success rates drop to 35\% (just above the 30\% threshold) but still represent a significant degradation.
These results highlight the inherent trade-offs in automated oracle design between sensitivity and specificity.
To address this, \alias allows developers to configure custom thresholds, thereby refining the oracle's precision over time.

\begin{figure}[t]
    \centering
    \begin{subfigure}[b]{\linewidth}
        \centering
        \includegraphics[width=0.8\linewidth]{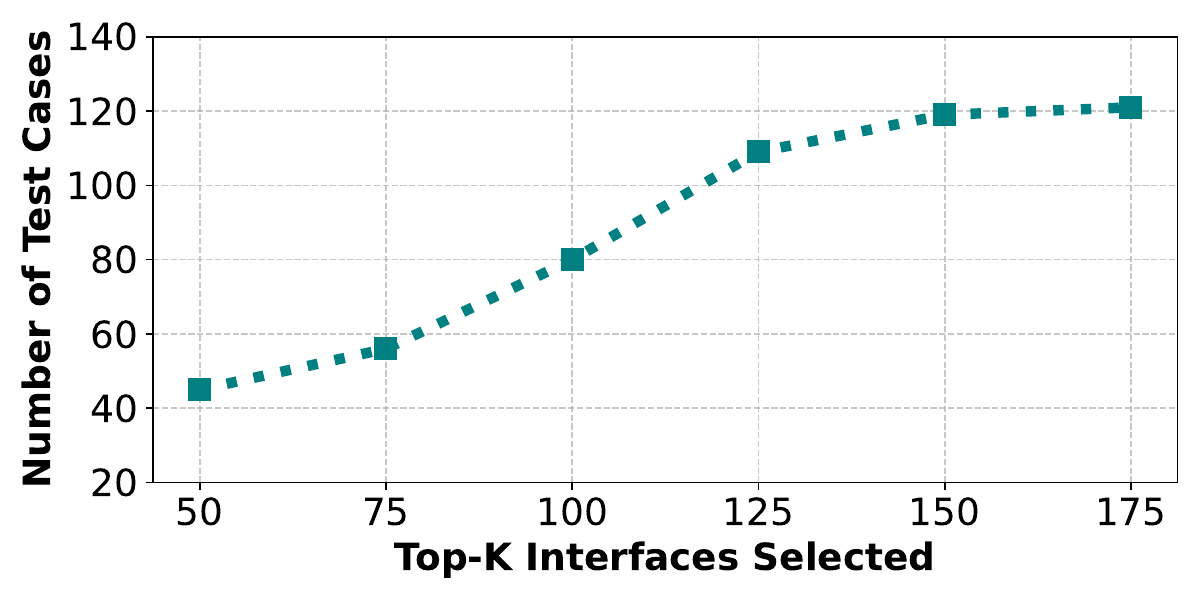}
        \caption{The Number of Test Cases vs. Top-K Interfaces Selected}
        \label{fig:test_cases}
    \end{subfigure}
    
    \vspace{0.2cm}
    
    \begin{subfigure}[b]{\linewidth}
        \centering
        \includegraphics[width=0.82\linewidth]{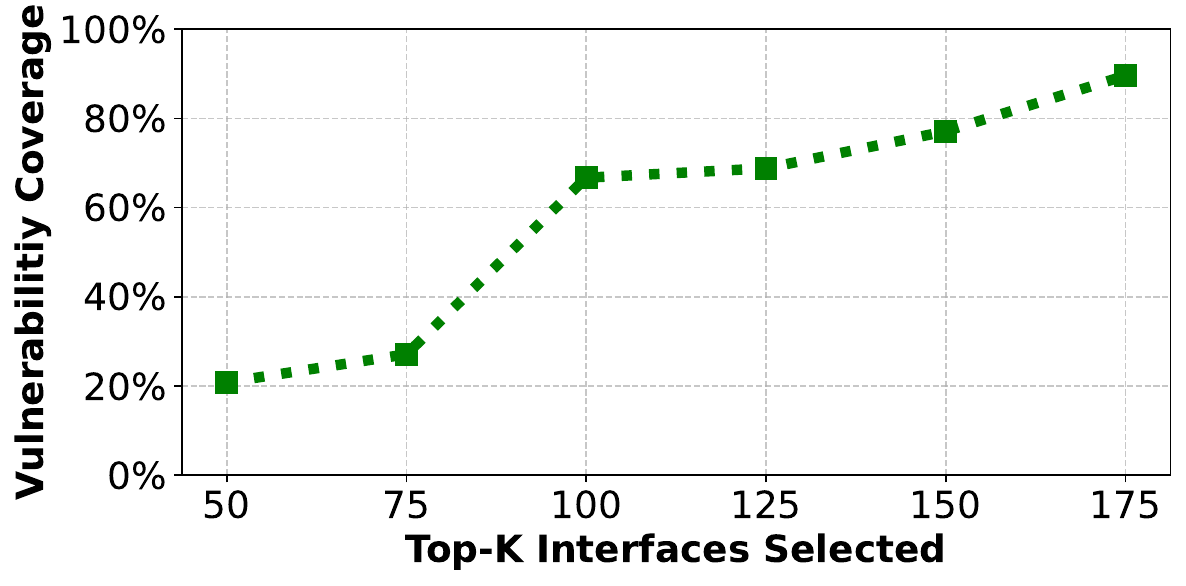}
        \caption{Vulnerability Coverage vs. Top-K Interfaces Selected}
        \label{fig:bug_coverage}
    \end{subfigure}
    
    \caption{Sensitivity Analysis with Varying Top-K Interfaces}
    \label{fig:sensitivity_analysis}
    \vspace{-9pt}
\end{figure}

We also analyze the 11 vulnerabilities that \alias fails to detect.
Besides the 3 false negatives, the remaining 8 bugs are due to test generation limitations where no test is ever created for the vulnerable paths. Particularly, 5 of these bugs are missed because the recorded traffic corpus does not contain requests exercising the vulnerable paths, a fundamental limitation of any record-and-replay-based approach. The other 3 are associated with interfaces whose complexity scores fall outside our top-K selection.
To mitigate this issue, \alias incorporates a historical execution-aware selection mechanism. This mechanism maintains a history of previously executed tests. When generating a new test suite, it prioritizes fault injections on interfaces that have not yet been tested or have previously failed, effectively avoiding the repetition of passed tests. This history is periodically reset (e.g., weekly) to ensure the system is re-validated against potential regressions, providing a balance between exploring new test space and continuous verification.

\subsection{Parameter Sensitivity Analysis}

To further investigate the trade-off between test suite size and bug coverage, we conduct a sensitivity analysis by varying the number of top-K interfaces from 50 to 175, as shown in Fig.~\ref{fig:sensitivity_analysis}.
We first examine the execution cost, which is quantified by the number of generated test cases (Fig.~\ref{fig:test_cases}).
The test suite grows substantially as the initial interfaces are added, since they introduce numerous unique endpoints. However, the curve begins to flatten as K increases, because the new traces are likely to contain endpoints that have already been covered by the previous traces.
This demonstrates the effectiveness of our pruning logic in eliminating redundant test cases.
Next, Fig.~\ref{fig:bug_coverage} shows the corresponding vulnerability coverage.
The curve shows a critical inflection point between K=75 and K=100, where coverage dramatically increases to nearly 70\%.
This validates our endpoint coverage-guided strategy for its ability to locate critical fault injection targets with sufficient traces.
Similar to Fig.~\ref{fig:bug_coverage}, this curve also shows diminishing returns, with each additional trace contributing progressively less to the overall coverage.


These two curves highlight the fundamental trade-off between coverage and cost. While a large K can achieve higher coverage in a single run, it demands the execution of a larger number of test cases, which may be impractical for frequent testing cycles. This is where \alias's historical execution-aware mechanism becomes critical. Instead of relying on a single, exhaustive run, teams can use a smaller K (e.g., 100) for daily or continuous testing. Over successive runs, \alias's history tracking ensures that it avoids re-testing previously passed scenarios and gradually explores other interfaces. This strategy allows \alias to achieve high cumulative coverage over time with smaller daily executions.

\subsection{Case Studies}
\label{sec:case_studies}

To further illustrate the types of bugs discovered by \alias and demonstrate the effectiveness of its design principles, we present two representative case studies from our deployments.

\begin{itemize}[noitemsep,leftmargin=5.5mm]
    \item \textbf{Case 1: Cascading Failure due to Unhandled HTTP Timeout.} A microservice, Service A, is responsible for processing uploaded data. Its workflow involves receiving a request, querying a third-party service via an HTTP REST API to validate and supplement the incoming information, and then publishing the final message to a Pulsar queue. 
    When \alias injects a \texttt{java.net.SocketTimeoutException} into this HTTP call, 
    the worker thread hangs indefinitely due to missing timeout policy.
    This causes the initial request to fail and prevents the thread from processing any subsequent requests, leading to a complete and persistent failure of that business function. The root cause is that the RestTemplate HTTP client used to call the third-party service is instantiated without a read timeout. This highlights the value of \alias's data-flow dependency prioritization. The workflow represents a classic Producer-Consumer pattern (User $\rightarrow$ Service A $\rightarrow$ 3rd Party Service $\rightarrow$ Pulsar), and \alias's complexity-driven selection prioritizes such multi-hop traces as they are more likely to cause cascading failures.

    \item \textbf{Case 2: Silent Failure in Asynchronous Communication.} A service responsible for adding new data entries first writes the data to a database and then publishes a notification event to a Kafka topic. 
    When \alias injects a \texttt{DisconnectException} 
    into the Kafka publish process, the service fails to capture this error. Consequently, the original REST API call still returns an HTTP 200 OK success code to the caller, despite the internal failure.
    The data is correctly persisted, but downstream services are never notified, leading to system-wide data inconsistency. 
    The root cause is a ``fire-and-forget'' pattern where the application triggers the asynchronous Kafka publish operation but fails to check its return value or handle exceptions. 
    This vulnerability underscores the critical need for \alias's granular assertion points. A naive oracle checking only the final HTTP response would miss this bug. By placing an assertion point directly at the internal Kafka producer endpoint, \alias correctly identifies the failure, revealing the silent but critical issue.
\end{itemize}

\subsection{Threats to Validity}

\textbf{External Validity.} Our evaluation is conducted solely on Huawei Cloud, which may limit its generalizability. However, we believe our findings are broadly applicable for the following reasons. First, the architecture patterns, platform components (Kafka, Redis, MySQL), and communication protocols (HTTP, gRPC) we target are 
widely adopted in industry.
Second, our design principles, i.e., black-box traffic analysis, application-level fault injection, and multi-faceted oracles, are platform-agnostic. Finally, the challenges we address (state dependencies, test space explosion, oracle design) are fundamental to resilience testing regardless of the specific cloud platform.

\textbf{Internal Validity.} Several factors may affect the accuracy of our results. First, reliance on threshold-based oracles may cause false positives or negatives. To mitigate this, \alias allows teams to customize thresholds based on their requirements and continuously refine them through operational feedback (Sec.~\ref{sec:phase4_test_oracle}). Second, our complexity-driven selection is heuristic and may miss bugs in simpler interfaces. We address this through our historical execution-aware mechanism that progressively explores untested interfaces over multiple runs, achieving cumulative coverage. Third, the record-and-replay approach is limited to execution paths present in captured traffic. While this is a fundamental constraint, we mitigate it by continuously recording new traffic patterns and maintaining a diverse trace corpus that grows over time, capturing an increasingly comprehensive set of execution scenarios.

\section{Lessons Learned}
\label{sec:lessons_learned}

The design, deployment, and evaluation of \alias in an industrial environment have yielded several critical insights into the nature of resilience testing for microservice systems. We distill our experience into the following key lessons.

\textbf{Verification Oracles Must Go Beyond the API Boundary.} Early in our work, we considered a simpler oracle model that only checked the final, user-facing API response (e.g., HTTP 200 OK), a common approach in existing research. This proved to be insufficient. We repeatedly encountered scenarios, particularly with asynchronous operations, where the primary API call would succeed while a critical background operation failed silently. As shown in our case study (Sec.~\ref{sec:case_studies}), this can lead to severe data inconsistencies that go undetected by simple oracles. This experience underscores a fundamental lesson: effective resilience testing requires deep, internal visibility. 
A trustworthy verification oracle must establish assertion points not only at the service boundary but also at critical internal endpoints, such as platform component interactions, to catch failures that do not immediately propagate to the user.

\textbf{Cumulative Coverage Through Automation Outweighs Perfect Prioritization.}
While many studies~\cite{DBLP:journals/tdsc/ChenCYLH24,DBLP:conf/icws/LongWCCCW20,DBLP:conf/sosp/LuL00TYY19,DBLP:conf/kbse/ChenDWQ20} leverage sophisticated strategies to prioritize fault injection points, we learned that in a complex, multi-team industrial setting, it is challenging to predict bug locations.
We found that the practical value of a testing system can be approximated by: \textit{Value = (per-run Coverage $\times$ Frequency)~/~Effort}.
A ``perfect'' prioritization strategy that requires substantial manual effort or complex system models may achieve high per-run coverage but low frequency due to operational overhead. In contrast, our ``good enough'' complexity-driven heuristic, combined with full automation, enables daily or weekly testing runs.
The key is to design a system that is history-aware to intelligently avoid re-testing passed scenarios. This allows the framework to accumulate effectiveness over time, systematically exploring new and lower-priority interfaces in successive runs.
This iterative approach achieves high cumulative coverage over time with smaller, more manageable daily executions, providing a more practical and robust path to improving system reliability.

\textbf{Engineering Trade-offs Are Inevitable in Industrial Systems.}
While academic prototypes often show promise on simple applications, industrial systems present challenges of complexity and scale that demand different engineering trade-offs. 
Throughout \alias's development, we faced numerous design decisions where theoretical optimality conflicted with practical constraints. For instance, our payload-based variable identification heuristic is less precise than static taint analysis but scales to hundreds of services without source code access. Our fixed oracle thresholds occasionally cause false positives but enable full automation without per-test manual configuration. Similarly, our complexity metric is simpler than graph-based algorithms but requires no system models or dependency maintenance. These trade-offs reflect a fundamental reality of industrial software engineering, namely, a deployed system that finds 80\% of bugs is more valuable than a perfect system that remains in the research lab. The key is to make these trade-offs consciously, provide configuration options for teams with specific needs, and continuously iterate based on operational feedback.

\section{Related Work}
\label{sec:related_work}

\subsection{Resilience Testing and Chaos Engineering}

Resilience testing through deliberate fault injection has evolved from coarse-grained infrastructure failures to sophisticated testing at the application level. Early approaches like Netflix's Chaos Monkey~\cite{bennett2012chaos,DBLP:journals/software/BasiriBRHKRR16} pioneered the practice of randomly terminating virtual machines to test system recovery. While effective for infrastructure robustness, these methods fail to exercise application-specific error-handling logic~\cite{DBLP:conf/icdcs/HeorhiadiRJRS16,DBLP:conf/cloud/MeiklejohnESMP21}. This limitation motivated finer-grained approaches. For example, Gremlin~\cite{DBLP:conf/icdcs/HeorhiadiRJRS16} intercepts inter-service messages, Filibuster~\cite{DBLP:conf/cloud/MeiklejohnESMP21} combines static analysis with concolic execution for HTTP services, while Doctor~\cite{10.5555/876900.881183}, Setsudo~\cite{DBLP:conf/sosp/JoshiGBGP13}, and FATE/DESTINI~\cite{DBLP:conf/nsdi/GunawiDJAHAASB10} inject lower-level I/O failures.

The core challenge in resilience testing is managing the combinatorial explosion of fault space~\cite{DBLP:conf/icws/LongWCCCW20,DBLP:journals/tse/WuYNNPHY23}. Sophisticated prioritization strategies have emerged to address this. For example, LDFI~\cite{DBLP:conf/sigmod/AlvaroRH15} reasons backward from correct executions to identify minimal fault sets, IntelliFT~\cite{DBLP:conf/icws/LongWCCCW20} uses fitness-guided feedback to steer toward impactful scenarios, and MicroFI~\cite{DBLP:journals/tdsc/ChenCYLH24} employs PageRank algorithms for prioritization. While theoretically elegant, these approaches face significant adoption barriers in industrial settings. LDFI requires deterministic execution models impractical for multi-team environments, fitness-guided methods need multiple iterations unsuitable for large-scale systems, and graph-based approaches struggle with dynamic service topologies. Recent surveys~\cite{DBLP:journals/software/PortocarreroCVR24} and quantum-ML optimizations~\cite{arxiv2506-02090} further highlight the complexity-efficiency trade-off. \alias addresses these practical constraints through a complexity-driven heuristic that requires no system models and converges in a single pass—trading theoretical optimality for operational feasibility, as our eight-month deployment validates~\cite{DBLP:conf/cloud/AlvaroASRBH16,DBLP:books/lib/Newman15,DBLP:journals/cloudcomp/TuckerHJBR18,DBLP:journals/tse/ZhangMHBM21}.

The evolution of testing frameworks reflects industrial needs. Specifically, CHESS~\cite{arxiv2303-07283} targets self-adaptive systems, adaptive chaos engineering uses reinforcement learning, Rainmaker~\cite{DBLP:conf/nsdi/Chen0NYX23} provides push-button testing for cloud applications and MicroRes~\cite{DBLP:conf/issta/YangLS0FYL24} profiles degradation patterns. Recent deployments reveal practical insights. For example, Chen et al.~\cite{arxiv2507-16109} demonstrated 80\% superior response stability in cloud-edge environments through extensive fault injection. The oracle problem remains critical. While LLM-based approaches like CANDOR~\cite{arxiv2506-02943} generate oracles through consensus, and SATORI~\cite{DBLP:conf/ase/AlonsoMLSBR25} derives them from API specifications~\cite{arxiv2411-01789,arxiv2505-07870}, \alias employs pragmatic threshold-based oracles with configurable parameters, enabling full automation without per-test manual configuration.

\subsection{Traffic Record-and-Replay and Mocking}

Creating realistic test scenarios through traffic record-and-replay has become essential for high-fidelity testing~\cite{DBLP:conf/kbse/Arora0IKR18,DBLP:conf/icse/LiuLDLZ22}. The fundamental challenge is state dependency, i.e., production traffic contains time-sensitive variables that cause replay failures. Web application tools like WaRR~\cite{DBLP:conf/dsn/AndricaC11}, Mugshot~\cite{DBLP:conf/nsdi/MickensEH10}, and WebRR~\cite{DBLP:conf/sigsoft/LongWC0020} address this through various browser-specific mechanisms. In microservices, the challenge is amplified due to distributed state across services. Liu et al.~\cite{DBLP:conf/icse/LiuLDLZ22} used static taint analysis for variable identification but require SOFA/SOFABoot framework coupling, while Zhu et al.~\cite{DBLP:conf/kbse/ZhuWWLCSZ20} leveraged AI for mocking point recommendations~\cite{DBLP:conf/kbse/ArcuriFG14,DBLP:conf/sigsoft/ArcuriFG15}.

Mocking, traditionally used in unit testing~\cite{DBLP:conf/msr/SpadiniABB17,DBLP:journals/ese/SpadiniABB19}, is increasingly integrated into system-level resilience testing. EvoMaster~\cite{DBLP:journals/tosem/ZhangALLX23} supports test seeding and service mocking, while MicroFuzz~\cite{DBLP:conf/icse/DiLG24} explicitly targets microservice environmental complexities. \alias distinguishes itself through a black-box, payload-centric approach that treats services as opaque entities. Our two-stage heuristic, i.e., identifying tokens appearing in both request and response, then filtering by cross-span variability, achieves 98.3\% replay success without source code access or framework dependencies. This design choice, while potentially less precise than white-box analysis~\cite{DBLP:conf/asplos/LiuLLLLGT17}, proves crucial for polyglot industrial environments where teams use diverse languages and frameworks. Unlike approaches requiring extensive system knowledge or training data, \alias provides immediate value through lightweight dynamic analysis, making it practical for the continuous evolution of production systems.

\section{Conclusion}
\label{sec:conclusion}

In this paper, we present \alias, a framework that addresses the critical challenges of resilience testing in production microservice systems. The core contribution of \alias is its holistic methodology for creating realistic, scalable, and fully automated resilience tests. It bridges the gap between testing and production reality by combining high-fidelity workloads, derived from live traffic, with a library of fine-grained, application-level faults. To navigate the combinatorial complexity, \alias provides a complexity-driven approach that prunes the vast test space, enabling an efficient yet comprehensive evaluation of the most critical execution paths. Deployed on Huawei Cloud for over eight months, \alias successfully uncovers 137 potential resilience vulnerabilities, with 89 confirmed by developers. Furthermore, we also perform controlled experiments on a benchmark set of 48 reproduced bugs, where \alias achieves a high bug coverage of 90\%.
These results validate that our approach not only achieves high replayability of production traffic but also efficiently discovers critical, real-world vulnerabilities. 

\begin{acks}
This work is supported by the National Natural Science Foundation of China (No. 62402536).
\end{acks}

\balance
\bibliographystyle{ACM-Reference-Format}
\bibliography{bibliography}

\end{document}